# On the application of the Method of Dimensionality Reduction to two-dimensional contacts

Valentin L. Popov

Technische Universität Berlin, 10623 Berlin, Germany

**Abstract**. The conventional formulation of the Method of Dimensionality Reduction (MDR) in contact mechanics is only applicable two "point contacts", that is to contacts of two unbounded three-dimensional bodies over final contact area. We analyze here if it is possible to find an at least approximate formulation of the MDR for "line contacts", that is contacts which contact area is unbounded in one direction. In this case the problem can be formulated as contact of two two-dimensional half-"spaces".

## 1 Introduction

The conventional formulation of the Method of Dimensionality Reduction in contact mechanics is based on the existence of some interrelation between the indentation depth $d$ and the contact radius $a$, (see e.g. [1],[2]). On the other hand, it is known that the indentation depth is not unambiguously defined in two-dimensional contacts ("line contacts") [3]. Thus, the procedure of the MDR cannot be directly transferred to two-dimensional contacts. However, the divergence leading to the formally infinitely large indentation depth in a true two-dimensional contact is relatively weak (logarithmic). This means that for finite bodies the indentation depth can be defined approximately (with logarithmic accuracy). In the present paper we derive the approximate formulation of the MDR for finite 2D contacts.

## 2 2D-indentation of a flat-ended punch

Consider a two-dimensional contact of a "flat-ended cylinder", that is a rigid rectangle with half-width $a$. If the normal force per unit length in the normal direction is $P$, then the pressure distribution is given by [3]:

$$p(x) = \frac{P}{\pi(a^2 - x^2)^{1/2}} \qquad (1)$$

and the displacement outside the contact region is equal to

$$u_z(x) = d - \frac{2P}{\pi E^*} \ln\left|\frac{x}{a} + \left(\frac{x^2}{a^2} - 1\right)^{1/2}\right|, \quad x > a. \qquad (2)$$

The problem of two-dimensional contacts is that it is not possible to unambiguously determine the indentation depth relative to a point in infinity as the function (2) diverges logarithmically in infinity. In the present paper, we will define the indentation depth relative to some point far from the contact region, at the coordinate $x = L$, where the displacement will be considered to be zero:

$$u_z(L) = d - \frac{2P}{\pi E^*} \ln\left|\frac{L}{a} + \left(\frac{L^2}{a^2} - 1\right)^{1/2}\right| = 0. \qquad (3)$$

In the following, we assume that $L \gg a$, then $\Lambda = \ln\left|\frac{L}{a} + \left(\frac{L^2}{a^2} - 1\right)^{1/2}\right| \approx \ln\left(\frac{2L}{a}\right)$. The quantity $\Lambda$ is the Function of both $L$ and $a$, but if we allow only relatively small changes of $a$ (say for one order of magnitude) then $\Lambda$ can be considered as a constant (with "logarithmic accuracy").

For thus defined indentation depth we get

$$d = \frac{2P}{\pi E^*}\Lambda. \qquad (4)$$

If the normal force is changing by an infinitesimally small amount of $\mathrm{d}P$, then the change in the indentation is equal to

$$\mathrm{d}d = \frac{2}{\pi E^*}\Lambda \mathrm{d}P. \qquad (5)$$

The change in the displacement field is given by

$$\mathrm{d}u_z(x) = \frac{2}{\pi E^*}\left[\Lambda - \ln\left|\frac{x}{a} + \left(\frac{x^2}{a^2} - 1\right)^{1/2}\right|\right]\mathrm{d}P = \left[1 - \frac{1}{\Lambda}\ln\left|\frac{x}{a} + \left(\frac{x^2}{a^2} - 1\right)^{1/2}\right|\right]\mathrm{d}d, \qquad (6)$$

and the change in the pressure is equal to

$$p(x) = \frac{\mathrm{d}P}{\pi(a^2 - x^2)^{1/2}} = \frac{E^*}{2(a^2 - x^2)^{1/2}} \cdot \frac{1}{\Lambda}\mathrm{d}d \qquad (7)$$

Now we consider a contact between a rigid indenter with shape $z = f(x)$ and an elastic half-space, which is characterized by the effective elastic coefficient $E^*$:

$$E^* = \frac{E}{1 - \nu^2}, \qquad (8)$$

where $E$ is Young's modulus and $\nu$ Poisson-number.

The indentation depth under the effect of the normal line force $P$ is denoted as $d$ and the contact half-width as $a$. For a given profile of the indenter, each of these three variables can be determined by the other two; especially the indentation depth is a unique function of the contact radius, which we denote by

$$d = g(a). \qquad (9)$$

We consider the indentation process from the first contact to the final indentation depth and investigate the current values of the force, indentation depth and contact half-width $\tilde{P}$, $\tilde{d}$ and $\tilde{a}$. The entire process consists of a change in indentation depth from $\tilde{d} = 0$ to $\tilde{d} = d$, where the contact half-width changes from $\tilde{a} = 0$ to $\tilde{a} = a$ and contact force from $\tilde{P} = 0$ to $\tilde{P} = P$. The normal force at the end of the process can be written as follows

$$P = \int_0^P \mathrm{d}\tilde{P} = \int_0^a \frac{\mathrm{d}\tilde{P}}{\mathrm{d}\tilde{d}}\frac{\mathrm{d}\tilde{d}}{\mathrm{d}\tilde{a}}\mathrm{d}\tilde{a}. \qquad (10)$$

Taking into account that the stiffness of contact area with the half-width $\tilde{a}$ is given by

$$\frac{\mathrm{d}\tilde{P}}{\mathrm{d}\tilde{d}} = \frac{\pi E^*}{2}\frac{1}{\Lambda} \qquad (11)$$

and using relation (9), we obtain

$$P = \int_0^a \frac{\pi E^*}{2\Lambda}\frac{\mathrm{d}g(\tilde{a})}{\mathrm{d}\tilde{a}}\mathrm{d}\tilde{a}. \qquad (12)$$

Integration now results in

$$P = \frac{\pi E^*}{2\Lambda} g(a). \quad (13)$$

Let us now turn to the calculation of the pressure distribution in the contact area. An infinitesimal indentation depth by $\mathrm{d}\tilde{d}$ of an area with half-width $\tilde{a}$ produces the following contribution to the pressure distribution:

$$\mathrm{d}p(x) = \frac{\mathrm{d}P}{\pi(\tilde{a}^2 - x^2)^{1/2}} = \frac{E^*}{2(\tilde{a}^2 - x^2)^{1/2}} \cdot \frac{1}{\Lambda} \mathrm{d}d. \quad (14)$$

The pressure distribution at the end of the indentation process is equal to the sum of the incremental pressure distributions:

$$p(x) = \int_x^a \frac{E^*}{2(\tilde{a}^2 - x^2)^{1/2}} \cdot \frac{1}{\Lambda} \frac{\mathrm{d}d}{\mathrm{d}\tilde{a}} \mathrm{d}\tilde{a} \quad (15)$$

or with consideration of notation (9)

$$p(x) = \frac{E^*}{2\Lambda} \int_x^a \frac{1}{(\tilde{a}^2 - x^2)^{1/2}} \frac{\mathrm{d}g(\tilde{a})}{\mathrm{d}\tilde{a}} \mathrm{d}\tilde{a}. \quad (16)$$

The function $g(a)$, (9), thus, unambiguously determines both the normal force and the pressure distribution (16). Therefore, the problem is reduced to simply determining the function $g(a)$, (9).

The function $d = g(a)$ can be determined as follows: The infinitesimal displacement of the surface at the point $r = a$ caused by an infinitesimal indentation depth $\mathrm{d}\tilde{d}$ of contact area with the radius $\tilde{a} < a$ is equal to

$$\mathrm{d}u_z(a) = \left[1 - \frac{1}{\Lambda} \ln\left(\frac{a}{\tilde{a}} + \left(\frac{a^2}{\tilde{a}^2} - 1\right)^{1/2}\right)\right] \mathrm{d}\tilde{d}. \quad (17)$$

The total sinking of the surface at the end of indentation process is, therefore, equal to

$$u_z(a) = \int_0^d \left[1 - \frac{1}{\Lambda} \ln\left(\frac{a}{\tilde{a}} + \left(\frac{a^2}{\tilde{a}^2} - 1\right)^{1/2}\right)\right] \mathrm{d}\tilde{d} = \int_0^a \left[1 - \frac{1}{\Lambda} \ln\left(\frac{a}{\tilde{a}} + \left(\frac{a^2}{\tilde{a}^2} - 1\right)^{1/2}\right)\right] \frac{\mathrm{d}\tilde{d}}{\mathrm{d}\tilde{a}} \mathrm{d}\tilde{a} \quad (18)$$

or considering the notation (9)

$$u_z(a) = \int_0^a \left[1 - \frac{1}{\Lambda} \ln\left(\frac{a}{\tilde{a}} + \left(\frac{a^2}{\tilde{a}^2} - 1\right)^{1/2}\right)\right] \frac{\mathrm{d}g(\tilde{a})}{\mathrm{d}\tilde{a}} \mathrm{d}\tilde{a}. \quad (19)$$

This sinking, however, is obviously equal to $u_z(a) = d - f(a)$:

$$d - f(a) = \int_0^a \left[1 - \frac{1}{\Lambda} \ln\left(\frac{a}{\tilde{a}} + \left(\frac{a^2}{\tilde{a}^2} - 1\right)^{1/2}\right)\right] \frac{\mathrm{d}g(\tilde{a})}{\mathrm{d}\tilde{a}} \mathrm{d}\tilde{a}. \quad (20)$$

Partial integration and consideration of (20) leads to the equation

$$d - f(a) = \int_0^a \left[1 - \frac{1}{\Lambda} \ln\left|\frac{a}{\tilde{a}} + \left(\frac{a^2}{\tilde{a}^2} - 1\right)^{1/2}\right|\right] \frac{\mathrm{d}g(\tilde{a})}{\mathrm{d}\tilde{a}} \mathrm{d}\tilde{a} \quad (21)$$

$$= g(a) - \frac{1}{\Lambda} \int_0^a \frac{a}{\tilde{a}\sqrt{a^2 - \tilde{a}^2}} g(\tilde{a}) \mathrm{d}\tilde{a}$$

or

$$f(a) = \frac{a}{\Lambda} \int_0^a \frac{g(\tilde{a})}{\tilde{a}\sqrt{a^2 - \tilde{a}^2}} \mathrm{d}\tilde{a}. \quad (22)$$

The reverse transformation reads [4]

$$g(a) = \frac{2\Lambda}{\pi} a^2 \int_0^a \frac{\left(f(\tilde{a})/\tilde{a}\right)'}{\sqrt{a^2 - \tilde{a}^2}} \mathrm{d}\tilde{a} \quad (23)$$

With the determination of the function $g(a)$ the contact problem is completely solved.

The normal force can be easily received in the general form by inserting (23) into (13) which gives

$$P = E^* a^2 \int_0^a \frac{\left(f(\tilde{a})/\tilde{a}\right)'}{\sqrt{a^2 - \tilde{a}^2}} \mathrm{d}\tilde{a} \quad (24)$$

which coincides with the exact solution given in [6].

## 3 Examples

Let us consider as example a parabolic profile:

$$f(x) = \frac{x^2}{2R}. \quad (25)$$

Substitution into (23) gives

$$g(a) = \Lambda \frac{a^2}{2R}. \quad (26)$$

The normal force by Eq. (13)

$$P = \frac{\pi E^* a^2}{4R}. \quad (27)$$

For the pressure distribution we get with (16)

$$p(x) = \frac{E^*}{2R} \int_x^a \frac{\tilde{a}\mathrm{d}\tilde{a}}{(\tilde{a}^2 - x^2)^{1/2}} = \frac{E^*}{2R} \left(a^2 - x^2\right)^{1/2}. \quad (28)$$

Both (27) and (28) reproduce the well-known solution of this problem given by Landau and Lifshitz [5].

As a second example, let us consider a parabolic profile with a flat (produced e.g. due to wear), . The three-dimensional profile can be written as

$$f(x) = \begin{cases} 0, & x < b \\ \dfrac{x^2 - b^2}{2R}, & x > b \end{cases}. \qquad (29)$$

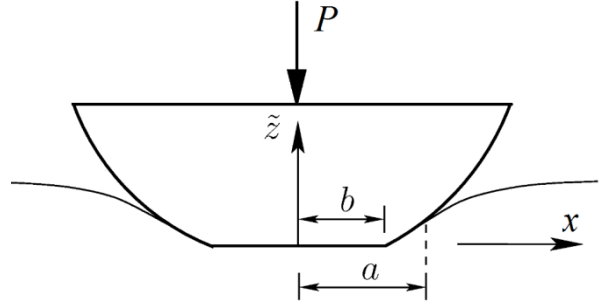

**Fig. 1** A cylindrical profile with a small flat of half-width $b$ in contact with an elastic half-space.

For the effective profile $g(x)$ we get according to (23)

$$g(a) = \frac{2\Lambda}{\pi} a^2 \int_0^a \frac{(f(x)/x)'}{\sqrt{a^2 - x^2}} \, dx. \qquad (30)$$

Substituting

$$\left(\frac{f(x)}{x}\right)' = \begin{cases} 0 & \text{für } 0 \leq x < b \\ \dfrac{1}{2R} + \dfrac{b^2}{2Rx^2} & \text{für } b \leq x \leq a \end{cases} \qquad (31)$$

into (30) results in

$$g(a) = \begin{cases} 0, & x < b \\ \dfrac{\Lambda}{\pi} \dfrac{a^2}{R} \int_b^a \left(1 + \dfrac{b^2}{\tilde{a}^2}\right) \dfrac{d\tilde{a}}{\sqrt{a^2 - \tilde{a}^2}} = \dfrac{\Lambda}{\pi} \dfrac{a^2}{R} \left[\dfrac{\pi}{2} - \arcsin\left(\dfrac{b}{a}\right) + \dfrac{b}{a}\sqrt{1 - \left(\dfrac{b}{a}\right)^2}\right], & x > b \end{cases}. \qquad (32)$$

To find the pressure distribution we need the derivative

$$\frac{dg(x)}{dx} = \frac{\Lambda}{R} \begin{cases} 0, & x < b \\ x\left(\pi - 2\arcsin\left(\dfrac{b}{x}\right) + 2\dfrac{b}{x}\right), & x > b \end{cases} \qquad (33)$$

For the pressure distribution we easily obtain

$$p(x) = \frac{E^*}{2R} \begin{cases} \displaystyle\int_b^a \frac{1}{(\tilde{a}^2 - x^2)^{1/2}} \tilde{a}\left(\pi - 2\arcsin\left(\dfrac{b}{\tilde{a}}\right) + 2\dfrac{b}{\tilde{a}}\right) d\tilde{a}, & x < b \\ \displaystyle\int_x^a \frac{1}{(\tilde{a}^2 - x^2)^{1/2}} \tilde{a}\left(\pi - 2\arcsin\left(\dfrac{b}{\tilde{a}}\right) + 2\dfrac{b}{\tilde{a}}\right) d\tilde{a}, & x > b \end{cases}. \qquad (34)$$

## 4 Discussion

We suggested a generalization of the Method of Dimensionality Reduction for the normal contact of two dimensional systems. As in the case of the conventional formulation of the MDR for three-dimensional contacts, it is based on the use of an auxiliary function. However, contrary to the 3d-formulation, a representation by a contact with an elastic foundation is not used.